# Missing Value Chain in Generative AI Governance
## China as an example

**Yulu Pi**
Centre for Interdisciplinary Methodologies
University of Warwick
yulu.pi@warwick.ac.uk

## Abstract

We examined the world's first regulation on Generative AI, China's Provisional Administrative Measures of Generative Artificial Intelligence Services, which came into effect in August 2023. Our assessment reveals that the Measures, while recognizing the technical advances of generative AI and seeking to govern its full life cycle, presents unclear distinctions regarding different roles in the value chain of Generative AI including upstream foundation model providers and downstream deployers. The lack of distinction and clear legal status between different players in the AI value chain can have profound consequences. It can lead to ambiguity in accountability, potentially undermining the governance and overall success of AI services.

## 1 Introduction

After the launch of ChatGPT, there has been a notable surge in public, regulatory, and academic discussions surrounding Generative Artificial Intelligence (Generative AI). This rapidly advancing branch of AI specializes in producing open-ended content, ranging from generating text [15] and code [8], to crafting vivid images [12] and videos [16], and even orchestrating melodies [7]. The value chain of Generative AI is layered [11]: it begins with foundation models like OpenAI's GPT-4, which can generate human-like text. These foundation models can then be fine-tuned and adapted for more specific applications, such as Duolingo Max [1] for language learning and Khan Academy's Khanmigo [19] for AI-assisted tutoring.

On April 11 Chinese regulators issued draft Measures to govern Generative AI service provisions in China. After seeking public feedback, the Cyberspace Administration of China (CAC) together with six other authorities finalized and announced The Provisional Administrative Measures of Generative Artificial Intelligence Services (生成式人工智能服务管理暂行办法, Generative AI Measures) with a total of 24 articles, on 13 July 2023, which are slated to become effective from 15 August 2023 [13]. It is the world's first effective regulation on Generative AI.

Through our analysis, we found that Generative AI Measures reflects the technical developments of generative AI, trying to provide a holistic governance framework that spans the entire life cycle of Generative AI. However, it provides an ambiguous differentiation between the various actors in the generative AI value chain, falling short in assigning appropriate responsibilities related to "algorithm design, the selection of training data, model generation and optimization, and the provision of services[1]" to the corresponding entities.

---

[1] The Generative AI Measures, Article 4,(2)

## 2 The Chinese Governance approach for Generative AI

The Generative AI Measures is predominantly designed for Generative AI services to the general public in China for the generation of texts, images, audio, videos, or other content. Consequently, any organization or individual utilizing generative AI technology to render such services, whether it's through online platforms or programmable interfaces, is subjected to this governance framework. Notably, sectors like scientific research and industrial applications, initially considered in the draft measures [14], have been exempted from the final regulations, underscoring China's dedication to promoting AI research and innovation. The Generative AI Measures strives to provide a holistic governance approach which covers the design and development process of the Generative AI services, algorithmic filling, criteria for content generation, the post-market monitoring and oversight. We have identified 6 emerging themes from the Generative AI measures, these include: **Data governance, Model governance, Algorithmic Filling, Content Governance, User Protection Governance as well as Departments in Charge.**

Since the training data has a direct impact on the generated outcomes, it naturally becomes a subject of regulation and is accompanied by compliance requirements. The Generative AI Measures emphasizes the lawful sources of training data and mandates effective measures to increase the truthfulness, accuracy, objectivity, and diversity. Article 8 specifically lays out requirements for the data annotation. Notably, the mandates concerning data governance were changed from obligatory requirements in the draft Measures to guidelines in the finalized Generative AI Measures. This shift not only underscores the significance of data governance of generative AI but also reduces immediate compliance burdens on enterprises, aligning with the current state of industrial practice.

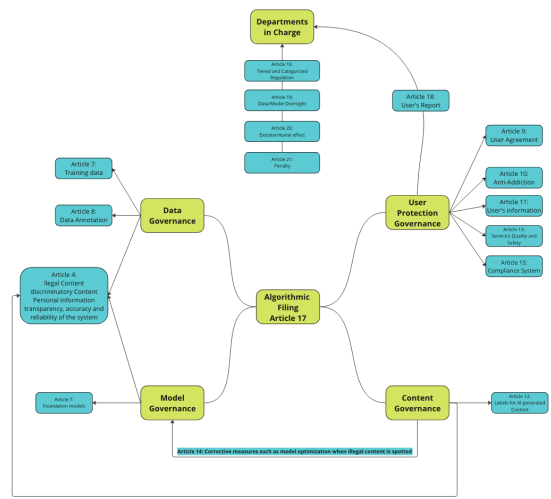

Figure 1: 6 emerging themes from the Generative AI measures

The Generative AI Measures introduces the notion of "foundation models" for the first time. This reflects the value chain of Generative AI, ranging from "foundation models" to "fine-tuned models" and then "service applications". Article 7 mandates the use of foundation models from legitimate sources. The safety assessment obligations and record-filing procedures with China's cyberspace agency required by Article 17 might provide a legal basis for determining legitimate sources. For instance, foundation models included in the deep synthesis service algorithm records [3] can be considered to have a legitimate source. For instance, following a filing application and approval by the Chinese government, Ernie Bot, Baidu's generative AI chatbot, is now fully accessible to the public in September 2023 [5]. Under Article 20, the CAC and relevant departments have the authority to direct extraterritorial generative AI service providers operating within China to adopt technical or other essential adjustments if they aren't adhering to Chinese laws. Although this suggests that those abiding by the law can still offer their generative AI services in China, currently, no foreign entities are active in this vast market. Moreover, the legitimacy of models trained outside of China, like the widely-used open-source model LLaMA, remains to be further clarified.

Article 14 integrates content governance with model governance, emphasizing that AI oversight needs to adapt to the continuously evolving nature of Generative AI. It states that providers must take immediate action upon identifying illegal content. This involves stopping its generation and dissemination, removing the content, undertaking corrective actions to improve the models, and promptly notifying the appropriate authorities. In addition to prohibiting illegal and discriminatory content, Article 12 mandates that providers must label AI-generated content. This ensures that users are aware they are engaging with content generated by AI, thereby enabling rational filtering and value judgment of the relevant content. The regulatory scope of the Generative AI Measures falls on the Generative AI services provided to the Chinese public, emphasizing the services' public-facing components. Numerous provisions specifically aim to protect user rights, requiring Generative AI service providers to uphold cybersecurity obligations as online information content producers. Moreover, as personal information handlers, these service providers need to comply with personal information protection obligations. These stipulations mandate the signing of a user agreement (Article 9), implementing measures to deter excessive dependence or addiction to the service (Article 10),, limiting acquisition to users' personal data and records (Article 11), ensuring a stable and secure service (Article 13), and setting up a system for complaints (Article 15).

The Generative AI Measures has inherited the power of algorithm service inspection from the *Provisions on the Administration of Algorithm-generated Recommendations for Internet Information Services* (互联网信息服务算法推荐管理规定) and *Administrative Provisions on Deep Synthesis in Internet-based Information Services* (互联网信息服务深度合成管理规定). Based on provisions like algorithmic filing and algorithm safety assessment, the Generative AI Measures establishes a mechanism for algorithm disclosure. Article 19 stipulates that providers must fulfill additional algorithm disclosure duties to the regulatory authorities. Relevant departments can require Generative AI service providers to supply necessary information that might affect user trust and choices. These details include 'sources, models, types, tagging rules, algorithm mechanisms'. Should any service contravene Chinese statutes, protocols, or the clauses of the articles, the CAC holds the authority to halt or revoke the online privileges of such foreign Generative AI service providers within China's jurisdiction.

## 3 The Missing Value Chain

On January 10, 2023, China's Administrative Provisions on Deep Synthesis in Internet-based Information Services (Deep Synthesis Provisions) entered into effect [3]. Within these provisions, distinctions were made between deep synthesis service (DSS) providers and DSS technical supporters. This distinction mirrors the legislative understanding of the Generative AI value chain, recognizing that the creation of foundational models and their integration into products and services can be handled by different entities. It represents a positive step towards governing AI products according to their distinctive characteristics. Notably, leading experts have underscored that "Many AI products are not produced by a single organisation, but involve a complex web of procurement, outsourcing, re-use of data from a variety of sources" [9]. However, this clear understanding of the AI value chain seems to have disappeared in the subsequent Generative AI Measures.

While the Generative AI Measures introduced the notion of "foundation models" for the first time, in line with special considerations for these models as noted in other legislative efforts, it neglects the difference between technical supporters and service providers. The general use of the term "providers" obscured the distinct roles of players within the generative AI value chain, spanning from "foundation model" to "fine-tuned model" and finally, "service application." This can also lead to conceptual discrepancies within the Generative AI Measures. For instance, companies that adopt OpenAI's API to integrate ChatGPT into their applications which offer services to the public, are classified as Generative AI service providers and are governed by the Generative AI Measures. According to Article 4, these companies are required to take effective measures during the algorithm design, training data selection, model development, optimization, and service provision processes to prevent discrimination. However, these companies may find it challenging to comply with this stipulation, as they lack access to both the training data and the design choices behind the adopted model.

The European Commission's proposed Draft AI Act from April 2021 did not directly address foundation models. This Act is based on the 'New Legislative Framework' which adopts a standard product safety perspective [4]. However, since its introduction, there has been growing criticism that the AI Act might be too static for practical use [2]. The Act adopted a risk-based approach, with prohibitions on systems with unacceptable risk, stringent mandatory regulation of high-risk systems, and minimal management of AI systems with limited risk. Any company identified as providing high-risk AI systems must comply with strict rules on data management, information provision and transparency, risk management, and human oversight. On the other hand, "low-risk" AI systems are only subject to meet minimal transparency commitments. Such a risk-based approach has been criticized for disregarding the complexity of use intention in determining the risk level. For instance, an image recognition algorithm might initially qualify as low-risk under the regulation. However, if it is later employed for facial recognition, the stringent responsibilities associated with high-risk AI would apply. This becomes more apparent in the case of foundation models, which can be adapted for a variety of uses. As it stands, the Act places the obligation on downstream deployers who adapt or fine-tune existing foundation models into specific use contexts. Consequently, the deployers would bear the more onerous duties and full responsibility for risks arising from external data and design decisions made during the foundation model's design stage, typically undertaken by large corporations.

In response to these criticisms as well as the rapidly evolving landscape of Generative AI, the Compromise Amendments have taken into account the complexity of the value chain for Generative AI systems, clarifying the legal situation of providers of foundation models as well as downstream deployers. It acknowledges that foundation models hold growing importance to many downstream applications and systems as they can be repurposed in countless downstream tasks. In the added Article 28, it outlined a total of nine obligations for providers of foundation models, with three being particularly noteworthy [2]. Firstly, risk identification is paramount. While upstream providers of a foundation model might find it challenging to foresee all its potential use cases and address its risks, they are nevertheless already aware of certain risks, like biases and lack of representation in the training dataset. Article 28 b(2a) would make it mandatory to identify and mitigate reasonably foreseeable risks throughout development with appropriate methods such as with the involvement of independent experts, as well as the documentation of remaining non-mitigable risks after development. Secondly, the emphasis on testing and evaluation is vital. It ensures providers commit to adequate testing and evaluation to make design decisions that achieve appropriate levels of performance, predictability, interpretability, corrigibility, safety, and cybersecurity. Lastly, the providers of the foundation models also bear extremely extensive documentation obligations. This facilitates downstream providers in meeting their obligations as providers of high-risk AI systems. While the EU's approach to regulating Generative AI has been applauded as a step in the right direction, some critics argued that the associated compliance costs might be so high that only large companies can afford them. This could potentially stifle competition and hinder innovation [10].

As the first formally approved Generative AI service regulation on a global scale, the Generative AI Measures can gain valuable insights from the EU's approach to delineating different actors in the AI value chain, namely the upstream foundation model providers and downstream deployers. The responsibilities and obligations of upstream foundation model developers and downstream service providers need to be meticulously delineated, taking into account the AI production workflow, societal impacts, and economic benefits. The lack of distinction and clear legal status between different players in the AI value chain can have profound consequences. It can lead to ambiguity in accountability, potentially undermining the governance and overall success of AI services. If stringent governance disproportionately falls on SMEs and start-ups that adapt these foundational models for specific use scenarios, rather than major companies that own the foundational models with vast data access and computing power, it could impact the economy adversely [2].

## 4   Closing Remarks

As the world's first Generative artificial intelligence regulation, the Generative AI Measures provides valuable insights into data, model, and content governance. It also provides guidelines for user protection, helping other nations navigate the complexities of Generative AI oversight. We found that the Measures places a strong emphasis on user protection and content management, granting users the right to challenge or complain about AI systems. Moreover, Generative AI service providers are mandated to provide timely and transparent responses regarding the measures taken to address these concerns. Users are also empowered to report such issues to the relevant governing authorities. However, the Measures used the general term "providers" and assigned uniform responsibilities to various actors throughout the AI value chain, which can pose challenges in rule enforcement. In particular, deployers lack access to either model or training data or have technical resources to assess and alter the system when it malfunctions, violates regulations, or infringes users' rights, as stipulated by the Measures.

To address the ambiguities within the Generative AI Measures, a detailed understanding of how various actors fit within the value chain of the Generative AI is essential. To begin with, the broad term "provider" should be substituted with the more specific definitions of upstream model providers and downstream deployers. This involves comprehensively mapping their roles and responsibilities across stages, from data creation and curation to training, adaptation, and deployment [6]. Upstream foundation model developers might be required to adopt a limited, staged release [17]. By initially granting access only to security researchers and a select cohort of stakeholders, potential vulnerabilities or issues can be identified and resolved prior to a broader public release. In parallel, it's crucial for downstream deployers to enhance their mechanisms for responding to user feedback and complaints. An open and responsive communication system between foundation model providers, application deployers and application users is instrumental in identifying and mitigating potential misuse. This collaborative framework enables upstream developers to offer instructions and maintain vigilant oversight over how their models are being used to safeguard and update post-deployment. The general purpose of foundation model makes the risk evaluation and mitigation challenging as many harms are only discovered after a model has been integrated into downstream products and used by a large user base. Therefore, effective collaboration between all stakeholders enables downstream deployers to promptly report identified issues based on user engagement, facilitating timely updates and communication with upstream model developers.

The series of legislative actions targeting specific algorithmic services or applications underscores China's recognition of the significance of AI oversight and its commitment to an agile governance strategy. On June 6, 2023, the State Council introduced its Legislative Work Plan for 2023, which emphasized the need for a Draft Artificial Intelligence Law [18]. As of now, China has so far issued the most AI legislation, with a predominant focus on its corporate applications. We contend that forthcoming legislation should align the governance of AI with its value chain, ensuring that the oversight of generative AI harmonizes its entire production and deployment life cycle. Within the Generative AI Measures, Articles 3 and 16 distinctly emphasize "categorized and graded governance," underscoring the responsibilities of several domain-specific departments, such as internet information, education, broadcasting and television, and news and publishing. We envision a domain-specific, context-based governance framework that adeptly acknowledges the distinct roles of different actors throughout the AI value chain.